\def\AaA{{A\&A}}

\def\ApJ{{ApJ}}

\def\MN{{MNRAS}}
\def\Nat{{Nat}}
\def\PASJ{{PASJ}}

\def\spose#1{\hbox to 0pt{#1\hss}}
\def\approxlt{\mathrel{\spose{\lower 3pt\hbox{$\sim$}}
	\raise 2.0pt\hbox{$$<$$}}}
\def\approxgt{\mathrel{\spose{\lower 3pt\hbox{$\sim$}}
	\raise 2.0pt\hbox{$>$}}}

\def\multleft#1{\hbox to size{\vbox {\halign {\lft{##}\cr #1}}\hfill}\par}
\def\multright#1{\hbox to size{\vbox {\halign {\rt{##}\cr #1}}\hfill}\par}

\def\today{\ifcase\month\or January\or February\or March\or April\or May\or
      June\or July\or August\or September\or October\or November\or December\fi
      \space\number\day, \number\year}
\def\$<${\thinspace}

\def\boxit#1{\vbox{\hrule\hbox{\vrule\kern3pt\vbox{\kern3pt
          #1 \kern3pt}\kern3pt\vrule}\hrule}}

\def\keV{{\rm\thinspace keV}}

\documentstyle[epsfig]{mn}
\begin{document}

\hsize=6truein

\title[Line profiles from a disk around a Kerr black hole]{The profile and equivalent width of the X-ray iron emission-line
from a disk around a Kerr black hole}

\author[Dabrowski et al.]
{\parbox[]{6.in} {Y. Dabrowski$^1$, A.C. Fabian$^2$, K. Iwasawa$^2$, 
A.N. Lasenby$^1$, C.S. Reynolds$^2$\thanks{Present address: JILA,
University of Colorado, Boulder, CO 80302, USA}\\
\footnotesize
1. Mullard Radio Astronomy Observatory, Cavendish Laboratory, Madingley
Road, Cambridge CB3 OHE\\ 
2. Institute of Astronomy, Madingley Road, Cambridge CB3 0HA \\
}}                                            
\maketitle

\begin{abstract}
Recent X-ray observations have shown broad, skewed iron line emission
from Seyfert 1 galaxies which is explained by the emission being
fluorescence on a disk close to a black hole. During one interval, the
line in MCG--6-30-15 was so broad and redshifted that a Kerr black hole
is implied. We are therefore studying the effects of the Kerr metric on
the line profile, and extending the work by Laor and Kojima which dealt
only with extreme values of the spin parameter. Here we report that the
spin parameter of the black hole in MCG--6-30-15 is high ($a/M>0.94$),
and invert the line profile to obtain the disk emissivity profile, which
approximates a power-law. Continuum radiation returning to the disk
because of the Kerr metric does not enhance the equivalent width of the
line seen above 3 keV by more than about 20 per cent if the continuum
source corotates with the disk.
\end{abstract}


\section{INTRODUCTION} 

X-ray observations of Seyfert 1 galaxies made with ASCA have shown that the
iron emission lines discovered with GINGA (Pounds et al. 1990; Matsuoka et al.
1990) are broad (Fabian et al. 1994; Mushotzky et al. 1995; Reynolds 1997;
Nandra et al. 1997). In particular, the line in MCG--6-30-15 was found to be
both broad and skew from a 4.5 day long observation in 1994 (Tanaka et al.
1995). The line extends about 2 keV below the rest frame energy of the line
(6.4~keV) and only 0.3~keV above. The line profile is well fitted by that
expected from fluorescence of matter on the surface of an accretion disk less
than 20 gravitational radii (i.e. $20r_{\rm g}=20GM/c^2$) from a black hole
of mass $M$ (Tanaka et al. 1995; Fabian et al. 1995).
Much of the skewness of the line is explained by gravitational redshift,
and is the first clear evidence for the effects of strong gravity.

Recent work examining variability in the line profile during the 4.5 day
observation has revealed that it broadened still further during a deep minimum
in the light curve (Iwasawa et al. 1996). The red (i.e. low energy) wing of the
line extended further to the red and the blue (i.e.  high energy) wing
disappeared. This behaviour can be explained if much of the emission originates
from within $6r_{\rm g}$ (Iwasawa et al. 1996). Since a disk around a
non-spinning, Schwarzschild black hole does not extend within this region, it
was concluded that the black hole must be spinning rapidly. In this case the
Kerr metric applies and frame dragging causes the disk to extend inward of
$6r_{\rm g}$.

Light bending effects, in particular the return of some disk radiation to the
disk itself, can then become significant (see e.g. Cunningham 1976). In
Iwasawa et al. (1996) it was proposed (without detailed calculation)
that these effects could enhance the strength (equivalent
width) of the fluorescent line relative to to the continuum, emitted above the
disk, as required by the observations.

In this Letter we examine the line emission from a disk around a Kerr black
hole, from the point of view of profile and of equivalent width, for the
parameters relevant to MCG--6-30-15. Previously only the line profile of either
a Schwarzschild black hole (Fabian et al. 1989) or a maximal Kerr black hole
(Laor 1991; Kojima 1991) have been available for fitting. Here we explicitly
fit for the angular momentum parameter of the black hole, as well as the
inclination angle of the normal to the disk to the line of sight. In addition,
results of a free-form fit to the radial emissivity profile of the disk in the
fluorescent line are given, again for the first time.

\section{Diskline model using the Kerr metric}

The propagation of radiation from a disk around a Kerr black hole has already
been studied (e.g. Cunningham 1975, 1976) and applied to model the observed
profile of a line emission from the disk (Asaoka 1989; Kojima 1991; Laor 1991;
Bromley, Chen \& Miller 1997).
The main parameters to which the spectra are sensitive are the
inclination angle of the observer from the axis of symmetry $i$, the radial
emissivity profile of the disk $\varepsilon(r)$ and the
angular momentum parameter of the hole $a_*=a/M$, where $a$ is the angular
momentum and $M$ is the black hole mass. (We note the model cannot discriminate
$a$ and $M$ separately).  We shall here investigate the effect of changes in
$i$ and $a_*$, and show a derived emissivity profile across the
disk, at fixed $i$ and $a_*$.  

\subsection{Assumptions and method of calculation}
\label{sec:ass-calc}
\begin{itemize}
\item The black hole belongs to the Kerr family of solutions.
The metric used is in the Boyer--Lindquist form.  The gravity of the disk
itself (and of the clouds responsible for the fluorescence) is negligible.

\item The accretion disk is assumed to be geometrically thin
(i.e. thickness $\ll$ smallest radius) and axially symmetric around the hole
axis of rotation.

\item The matter responsible for the fluorescent line is supposed
to lie on the surface of the disk and rotates on equatorial circular
geodesics. Emission starts at radius $r_{in}$, which is fixed to the radius of
marginal stability $r_{ms}$ (Novikov \& Thorne 1973), up to an arbitrary outer
radius $r_{out}$. No significant emission is expected
far away from the centre since the line emissivity, described
below, decreases rapidly (0.4\% and 0.002\% of its maximum for radii of
$15r_{\rm g}$ and $100r_{\rm g}$ respectively if $a_*=0.998$).
Previous studies (Tanaka et al. 1995) suggest an outer radius of
$\sim 10r_{\rm g}$--$16r_{\rm g}$, so that we decided to fix this parameter
to $r_{out}=15r_{\rm g}$.
This constraint might have to be relaxed in future studies.

\item Except for the case where we explicitly
fit for the emissivity profile (see Section~\ref{sec:emss-fit} below), then in
the rest frame of the emitting material (which we usually call the {\em disk
frame}), the emissivity of the fluorescent component $\varepsilon(r)$ [defined
as energy emitted per unit proper time per unit proper area] is assumed to
follow the law described in equation (11b) of Page \& Thorne (1974).  This law
is derived for the continuum emission of the disk, but we suppose that the line
flux (integrated across frequency in the emission rest frame) also follows this
law.  Since $\varepsilon(r)$ so defined vanishes at both $r_{ms}$ and
$r=\infty$, we expect this emissivity profile to be more realistic than a
simple power law model (Fabian et al. 1989).

\item We assume that the emitted fluorescent iron $K_{\alpha}$ line can be
approximated by a $\delta$-function in frequency, so that each emitted photon
has an energy $h\nu_o$ of 6.40~keV in the disk frame (Iwasawa et al. 1996). The
emission is assumed to be locally isotropic. Thus, in the disk frame, the
emitted intensity $I_e(r,\nu_e)$ and flux $F_e(r,\nu_e)$ are described by:
\begin{equation}
\label{eq:Fe}
F_e(r,\nu_e)=\pi I_e(r,\nu_e)=\varepsilon(r)\delta(\nu_e-\nu_o).
\end{equation}
\item The fluorescent photons are free to reach the observer
unless they cross the hole event horizon (lost radiation) or they return to the
disk itself. (See Section~\ref{sec-ret} and Cunningham 1976).  The observer is
assumed to be located in the asymptotically flat region of space-time.
In practice we use a distance of $1000r_{\rm g}$.
\end{itemize}

\begin{figure}
 \begin{center}
  \leavevmode
   \psfig{file=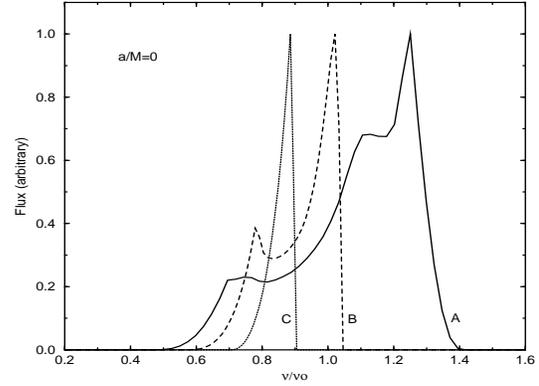, width=7cm, height=5cm}
   \caption{Predicted spectral line shapes for a Schwarzschild black hole
  ($a_*=0$).
	A is for  $i=85^o$, B for $i=30^o$,
            and C for $i=0^o$. \label{fig:profile_0}}
 \end{center}
\end{figure}

\begin{figure}
 \begin{center}
  \leavevmode
   \psfig{file=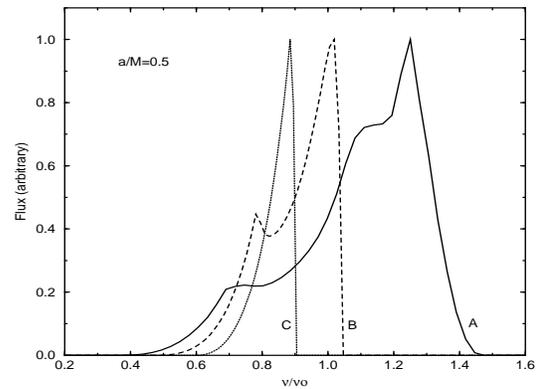, width=7cm, height=5cm}
   \caption{Same as Fig.~\ref{fig:profile_0} but for $a_*=0.5$. \label{fig:profile_0_5} }
 \end{center}
\end{figure} 

\begin{figure}
 \begin{center}
  \leavevmode
   \psfig{file=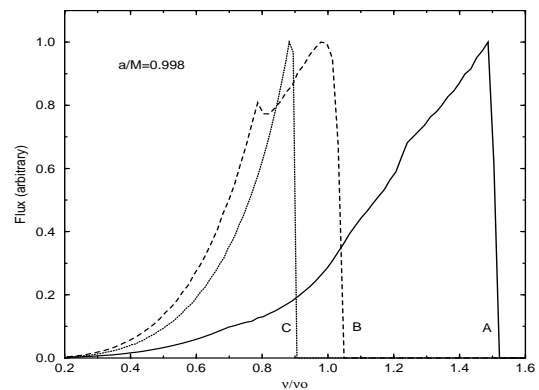, width=7cm, height=5cm}
   \caption{Same as for Fig.~\ref{fig:profile_0}, but for $a_*=0.998$,
  corresponding to an extreme Kerr black hole (Thorne, 1974). \label{fig:profile_1} }
 \end{center}
\end{figure} 

\begin{figure}
 \begin{center}
  \leavevmode
   \psfig{file=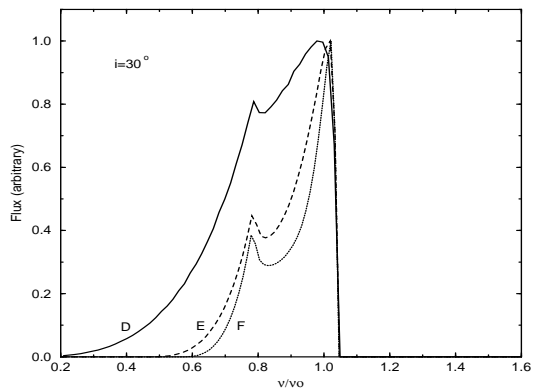, width=7cm, height=5cm}
   \caption{Predicted spectral line shapes at a fixed inclination $i=30^o$.
D is for an extreme Kerr black hole ($a_*=0.998$), E for $a_*=0.5$, and F
for a Schwarzschild black hole ($a_*=0$).\label{fig:profile_inc} }
 \end{center}
\end{figure}

The model then carries out a numerical integration of photon null geodesics
backwards in time, from the observer to the disk, taking into account all the
relativistic effects on both trajectories and redshift.  The photons are
emitted from the observer through a grid of equal solid angle pixels, which
could be those of a CCD detector. At the end of the integration process, each
image pixel is then associated with an emission radius $r_e$ on the disk and a
redshift $z$. We may compute the flux at the observer by using the phase space
occupation number $I_{\nu}/\nu^3$, which is invariant along the entire
trajectory of the photon. Considering an interval $\nu$ to $\nu+\Delta\nu$ in
observed frequency, then the observed flux in a pixel corresponding to solid
angle $\Delta\Omega$ at the observer, will be
\begin{equation}
F(r_e,z)=\frac{\Delta\Omega}{\pi\, \Delta\nu}\frac{\varepsilon(r_e)}{(1+z)^4},
\end{equation}
if the interval $(\nu,\nu+\Delta\nu)$ contains $\nu_0/(1+z)$, and zero
otherwise. Since redshift and flux are known for each pixel, images of
those quantities can be plotted, such as those in Luminet (1979) or
Bromley et al. (1997), but covering a wider range of parameters. Binning
all $F(r_e,z)$ into the redshift axis, enables one to compute the
observed line profile as seen by a distant observer.

\subsection{The line profile as a function of spin parameter and inclination}

Figs.~\ref{fig:profile_0}, \ref{fig:profile_0_5}, \ref{fig:profile_1} and
\ref{fig:profile_inc} show predicted spectral lines over a range of $a_*$
and $i$ parameters and suggest that, for fixed emissivity profile, the
overall line shape, in particular its width, is most sensitive to the
inclination angle.
For sufficiently high $i$ (i.e. $i>25^o$) the spectrum
is double-peaked. The low redshift component is mainly due to the receding
part of the disk from the observer and the high redshift photons essentially
come from the approaching half-part, as shown below.  In the case of more
face--on inclinations (i.e. $i<25^o$), the radial Doppler
effect is not efficient enough and the observed line shape is a one-peak line
with a weak blue component and a rather spread red contribution.  Over the
whole range of $i$, the red wing of the spectrum is being stretched
by transverse Doppler and gravitational redshifts.  This effect is boosted as
the angular momentum $a$ increases since the inner part of the disk then lies
closer to the hole.  Thus for a given $i$, the angular momentum
controls the red wing extension (see Fig.~\ref{fig:profile_inc}). 
For a given $a_*$, $i$ mainly constrains the upper limit of the blue part of
the line (see Figs.~\ref{fig:profile_0}, \ref{fig:profile_0_5} and
\ref{fig:profile_1}).

\begin{figure}
 \begin{center}
  \leavevmode
   \psfig{file=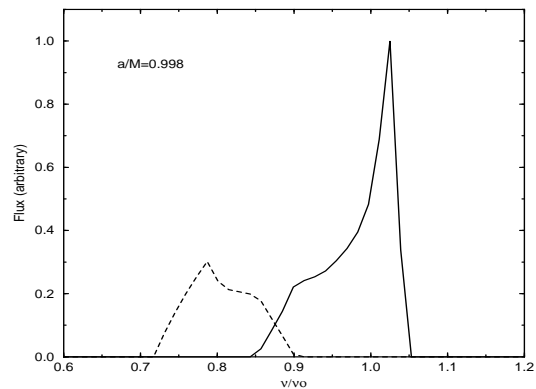, width=7cm, height=5cm}
   \caption{Contributions to the total lineshape from the receding (dashed)
  and approaching (solid) sections of the disk for $i=30^o$,
  $a_*=0.998$, $r_{in}=10r_{\rm g}$ and $r_{out}=15r_{\rm g}.$
 \label{fig:half}}
 \end{center}
\end{figure}

The radial Doppler shifts are illustrated in Fig.~\ref{fig:half}, where
receding and approaching halves of the disk contributions are separated.
An extreme black hole is used in Fig.~\ref{fig:half} but with an emission
area located between $r_{in}=10r_{\rm g}$ and $r_{out}=15r_{\rm g}$, in
order to show a case where the two components are clearly separated in
frequency. Such an appearance may be detectable if the X-ray emission is
from a few localized flares which move around on the disk (see e.g.
Iwasawa et al. 1996). Note the asymmetry of the two components, with the
blue peak being much more intense than the red one. This blue boosting
effect arises from the rather large redshift difference between the two
halves.

\section{Fits to the line profiles of MCG--6-30-15}

\begin{figure}
 \begin{center}
  \leavevmode
   \psfig{file=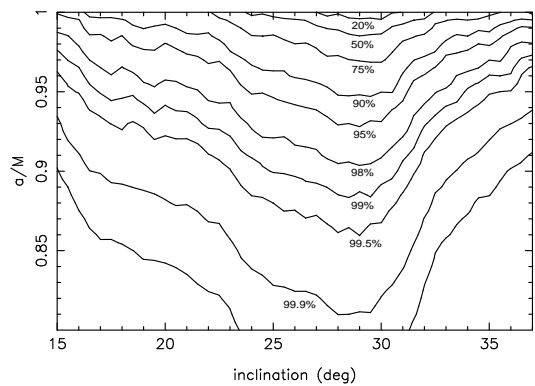, width=7cm, height=5cm}
   \caption{Contours of probability versus $i$ and $a_*$ are shown,
  calculated assuming a uniform prior on each parameter.   
 \label{fig:fit_contour}}
 \end{center}
\end{figure}

\begin{figure}
 \begin{center}
  \leavevmode
   \psfig{file=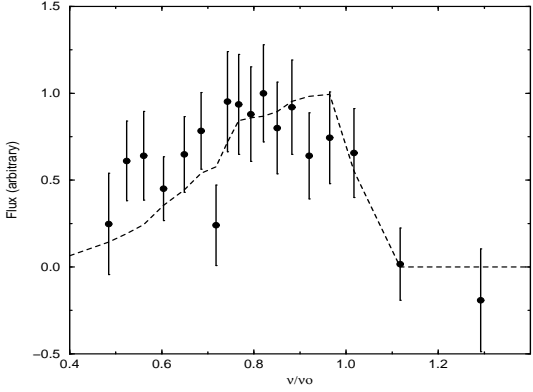, width=7cm, height=5cm}
   \caption{The line shape corresponding to the best-fit
values $i={29^{o}}^{\tiny +2.5}_{\tiny -3.2}$, $ 
a_*=1^{\tiny +0}_{\tiny -0.01}$
is compared to the data points from Iwasawa et al. (1996). The error bounds on
  the best fit parameters are 68\% confidence intervals obtained by
  marginalization with a uniform prior over the other parameter.}
 \label{fig:fit_ker}
 \end{center}
\end{figure}

The evidence of strong gravity around a Kerr hole in MCG--6--30--15 has already
been discussed in Iwasawa et al. (1996). This discussion included
a comparison between the
observed fluorescent $K_{\alpha}$ line at the period of lowest luminosity and a
predicted Kerr lineshape corresponding to $i=30^o$ and $a_*=0.998$
(Laor 1991).  We report here an actual fit using our diskline model in order
to obtain joint constraints on both $a_*$ and $i$.  The data consist
of 18 points lying between $\sim 3$\keV\ and $\sim 8$\keV\ and are described in
detail in Iwasawa et al. (1996).  We have carried out a $\chi^2$ fit assuming
the emissivity profile of Page \& Thorne as discussed above. 
The results are presented in Figs.~\ref{fig:fit_contour} and \ref{fig:fit_ker}.


The $\chi^2$ contours give relatively strong evidence for an inclination
angle of $\sim 25^o$--$30^o$ and strongly favour an extreme Kerr hole
($a_*> 0.94$) rather than a Schwarzschild hole ($a_*=0$). The inclination
$i \simeq 30^o$ arises mainly because of the observed cut--off for
energies higher than 6.5\keV, while the observed broad red wing
necessitates the high value of $a_*$. We note that a simple calculation
shows that significant line emission below 3.8 keV requires that the disk
extend within $1.9r_{\rm g}$ which in turn requires $a_*>0.95$.

\subsection{Fitting the emissivity profile}
\label{sec:emss-fit}

\begin{figure}
 \begin{center} 
  \leavevmode 
   \psfig{file=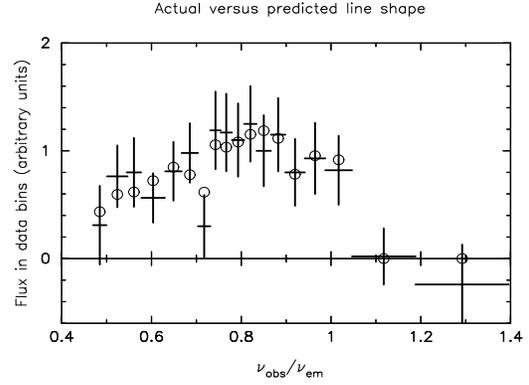,
   height=7cm,width=5cm,angle=-90} \caption{Comparison of actual data points (1
   sigma error bars shown) with the results in each bin predicted using
   the inferred emissivity profile (open circles).
   \label{fig:pred_bins_emss_fit}} \end{center}
\end{figure}

\begin{figure}
 \begin{center}
  \leavevmode
   \psfig{file=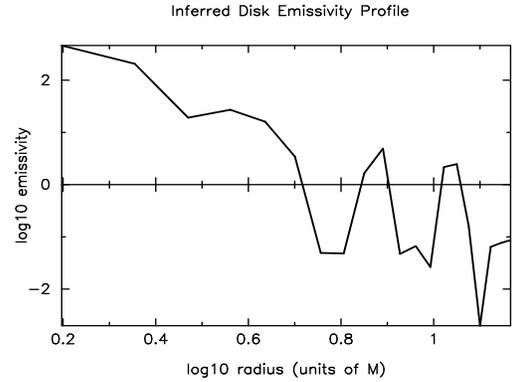, height=7cm,width=5cm,angle=-90}
   \caption{Log-log plot of inferred emissivity profile. \label{fig:emms_profile_emss_fit}}
 \end{center}
\end{figure}

The need for an $a_*$ value near 1 is so crucial in obtaining the highly
redshifted emission seen in the data, that the observed line can strongly
constrain the radial emissivity profile, $\varepsilon(r)$. This is because
while $a_*$ near 1 allows a much smaller inner radius of the disk, it is only
possible to take advantage of this if we weight the inner regions highly in
forming the line profile. This weighting function is of course basically the
emissivity profile. We have assumed above the Page \& Thorne form, which (for
the parameters of interest here) behaves approximately like $r^{-2.8}$ at large
$r$ and cuts off to zero at $r_{ms}$. It is of interest to try to establish
quantitatively the $\varepsilon(r)$ profile favoured by the data itself, to see
if this can reproduce something like a power law form, or even the cut off at
$r_{ms}$, favoured theoretically. As a first test case, one might attempt to do
this with the other important parameters, the inclination angle and the black
hole angular momentum parameter, held fixed. At a later stage one could then
try to include fitting of all three of $i$, $a_*$ and a free-format
$\varepsilon(r)$ simultaneously. We have written inversion software,
incorporating maximum entropy regularization, which can achieve this, and this
will be described fully, together with simulations, in a later paper. To show
results of immediate interest in relation to MCG--6--30--15, we show in
Figs.~\ref{fig:pred_bins_emss_fit} and \ref{fig:emms_profile_emss_fit} the
results obtained by this method without any regularization, using just
positivity of the emission as the only constraint, and with $i$ and
$a_*$ fixed at values plausible on other grounds of $30^o$ and 0.998
respectively. Such an approach is similar to the Lucy algorithm of Mannuci,
Salvati \& Stanga (1992) in application to inverting the profiles of
optical Balmer lines, except in the present case the integral kernel is
more complicated and has to be evaluated numerically.
Fig~\ref{fig:emms_profile_emss_fit} shows the inferred
emissivity profile as found in 20 radial bins, stretching from $r_{ms}$ to
$15r_{\rm g}$ in $r$.  Some evidence for a power law form is evident, with a
slope of perhaps $\sim -3.5$, although it obviously becomes noisy at the
smaller flux levels. Fig.~\ref{fig:pred_bins_emss_fit} compares the points
predicted using this radial profile (and assuming $i=30^o$ and
$a_*=0.998$) with the observed data. It is evident that a degree of overfitting
may be taking place, but that the results look sensible.


\section{The enhancement of equivalent width by returning radiation}
\label{sec-ret}
\begin{figure}
 \begin{center}
  \leavevmode
   \psfig{file=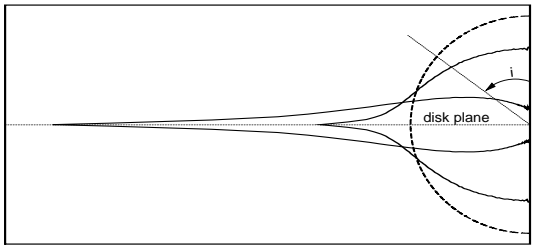, width=7cm, height=3cm}
   \caption{Received power as a function of emitted radius 
            (from left to right: $r=2r_{\rm g}$, $r=8r_{\rm g}$).
            The circular curve is for isotropic emission.
            \label{fig:polar}}
 \end{center}
\end{figure}

Light bending effects become very powerful for highly rotating holes
and should contribute to enhance the line equivalent width.
This bending effect is illustrated in Fig.~\ref{fig:polar} by a polar diagram showing
that for high inclination angles the emission strongly diverges from
an isotropic case.
The photons released from the inner part of the disk are preferentially
bent toward the plane of rotation.

\begin{figure}
 \begin{center} 
  \leavevmode
   \psfig{file=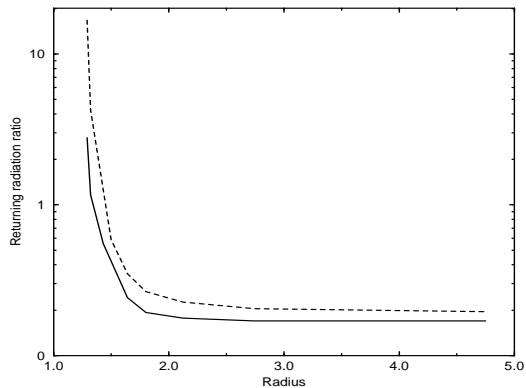, width=7cm,height=5cm}
   \caption{Returning radiation ratio :
$F_{ret}(r_{ret})/\varepsilon(r_{ret})$ (solid line). The dashed line
shows the ratio of returning photons over the number of emitted photons;
it is proportional to the enhancement in equivalent width (for a
continuum power-law of energy index 1). 
\label{fig:return_ratio}}
\end{center}
\end{figure}

\begin{figure}
 \begin{center}
  \leavevmode
   \psfig{file=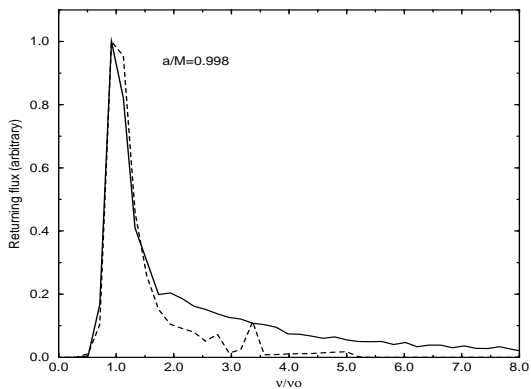, width=7cm,height=5cm}
   \caption{The line profile as seen by an observer rotating with the disk
at a radius of $1.5 r_{\rm g}$ (solid line), or $2 r_{\rm g}$
(dashed line). \label{fig:return_line}}
 \end{center}
\end{figure}

Moreover, a substantial number of the geodesics return to the disk itself ---
this is the returning radiation first treated by Cunningham (1976).  To study
the returning radiation, we used the same method of calculation as described in
Section~\ref{sec:ass-calc}, but placing the observer at a given radius
$r_{ret}$ on the rotating disk.  The result is shown in
Fig.~\ref{fig:return_ratio} as the ratio of returning flux $F_{ret}(r_{ret})$
(coming from any part of the disk) over emitted flux $\varepsilon(r_{ret})$ at
the same point (assuming again the Page \& Thorne law).  One can notice that
for inner radii (i.e. $r_{ret}<\sim 1.4r_{\rm g})$ this ratio is larger than unity.
Two effects from the returning radiation should be involved in enlarging the
equivalent-width of the fluorescent line relative to the primary power-law
continuum:

(i) More than half the continuum flux may hit the disk and produce
fluorescent photons.

(ii) Some of the returning photons are strongly blueshifted
(Fig.~\ref{fig:return_line}). This effect is particularly enhanced on the
inner parts of the disk. Then, since the power law slope of the continuum
is negative ($\sim -2$), at any given frequency, the continuum flux
incident on the disk is higher than the continuum flux which reaches the
observer.

Since most of the line emission within $r\sim 1.4r_{\rm g}$ emerges
below 2 keV, the line enhancement due to returning radiation has little
effect on the observed profile. An enhancement is instead needed
at $\sim 4$ keV, or from $r\sim 2r_{\rm g}$, where returning radiation
has little effect. Either the geometry of Martocchia \& Matt (1996),
in which the continuum source does not corotate with the disk, or some
other effect (e.g. high ionization of iron in the disk) is needed for
the observed increase in line strength (Iwasawa et al. 1996).

More detailed calculations of line strength in both the geometry considered in
this paper and that of Martocchia \& Matt will be presented in a later paper
(Lasenby et al. in preparation). This will cover a wider range of applications,
including predicted disk images and some differences from previous results
for line profiles from accretion disk will be highlighted.

\section*{ACKNOWLEDGEMENTS} ACF thanks the Royal Society for support.

\end{document}